\documentstyle[11pt,newpasp,twoside,epsf]{article}
\def\edcomment#1{\iffalse\marginpar{\raggedright\sl#1\/}\else\relax\fi} 
\reversemarginpar 

\begin{document} 
\title{Statistical Study of the Blue Straggler Properties 
in Galactic Globular Clusters}

\author{Francesca De Angeli and Giampaolo Piotto} 
\affil{Dipartimento di Astronomia, Universit\`a di Padova,
        Vicolo dell'Osservatorio, 2, I-35122 Padova, Italy}

\begin{abstract} 
In this paper we report on the most significant results from a
statistical analysis of the main properties of globular cluster blue
straggler stars (BSS) extracted from the HST snapshot database of
photometrically homogeneous CMDs (Piotto et al. 2002).  The BSS
relative frequency presents a significant anticorrelation with the
collisional rate and with the cluster total absolute luminosity.
\end{abstract}

\section{The Relative Frequency of Blue Stragglers} 
Our blue straggler star (BSS) catalog includes almost 3000 objects
extracted from 62 Galactic globular cluster (GC) CMDs which are part
of the photometrically homogeneous HST snapshot database (Piotto et
al. 2002).

In each cluster, the number of BSS has been normalized to the total 
population of stars. We defined three different specific frequencies
($F_{\rm BSS}$) as the ratios of the BSS counts to the number of HB,
RGB, and HB plus RGB stars, respectively.  The results described below
turned out to be independent from the adopted $F_{\rm BSS}$.

Figure 1 ({\it upper panel}) shows that there is a significant
anticorrelation between $F_{\rm BSS}$ and the absolute total magnitude
$M_V$ of the cluster.  In particular, the faintest clusters in our
sample (NGC~6717 and NGC~6838) present a BSS specific frequency that
is more than a factor of 20 larger than that in the brightest clusters.
Interestingly enough, PCC clusters ({\em open circles}), with the
exception of NGC 5946 that shows an anomalously low value of $F_{\rm BSS}$,
share the same distribution as the normal clusters.

A less significant dependence is found with the central cluster
density.  There is no clear correlation with the concentration
parameter.

King (2002) demonstrated
that the rate of stellar collisions in a King model GC is
$5\times10^{-15}(\Sigma_0^3r_c)^{1/2}$, where $\Sigma_0$ is the central
surface brightness in units of $L_{\odot V}$pc$^{-2}$ (equivalent to
$\mu_v=26.41$), and $r_c$ is the core radius in pc. As shown in Fig. 1 
({\it lower panel}), $F_{BSS}$ anticorrelates with this quantity.
The clusters with the highest collision rates have the smallest
relative number of BSS. The dispersion of the points in the {\it lower
panel} is larger than in the {\it upper panel}. We note a saturation in the
correlation for the clusters with collision rates smaller than
$\sim 10^{-9}$. The two clusters with the smallest total absolute
magnitude do not seem to partecipate to this general trend.


\begin{figure}[h!]
\plotfiddle{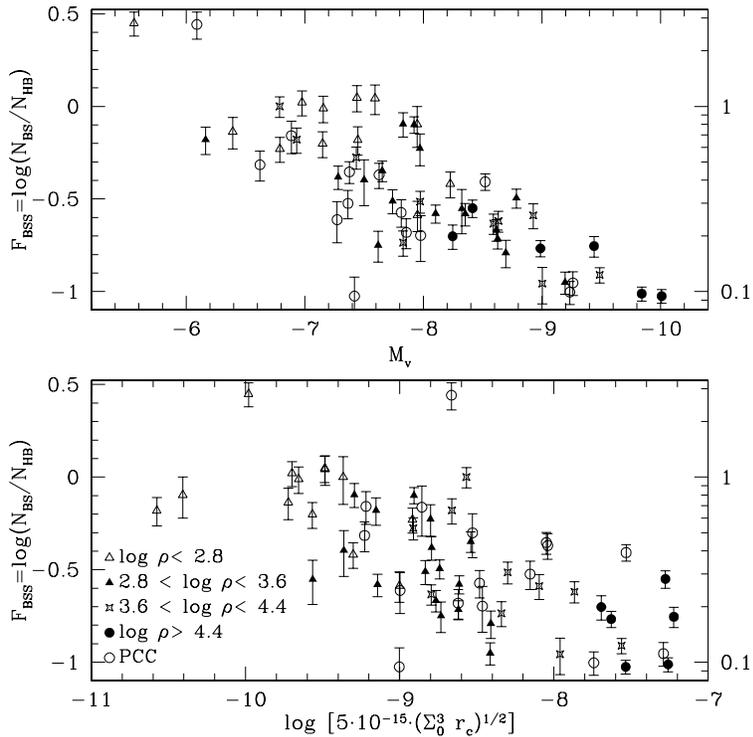}{8.9cm}{0}{49}{49}{-150}{-80}
\caption{The BSS relative frequency $F_{BSS}$ is plotted as 
a function of the total cluster magnitude ({\it upper panel}) and of
the collision rate ({\it lower panel})}
\end{figure}

\section{Conclusions}
The dependence of the BSS relative frequency with the cluster total
mass and with the expected collision rate is noteworthy.  Clearly, the
environment strongly affects the number of BSS that a GC can form at
the present time.  Stellar encounters can favour binary destruction:
separation of the soft binaries, merger of hard binaries.  The latter
might have given origin to a large number of BSS in the past, but now
high central density clusters have much less primordial binaries to
form BSS than low density objects.  A decrease in the number of
binaries means lower probability for a BSS to form.  Of course, high
density can also favour encounters, and therefore BSS formation via
collision. The anticorrelation between the binary frequency and the
collision rate of Fig.\ 1 suggests that direct collision of stars is
not the the dominant mechanism that triggers the formation of BSS, while
the number of surviving (primordial?)  binaries might be the relevant
parameter.

\end{document}